\documentclass[reprint,amsmath,amssymb,aps,pra,floatfix]{revtex4-2}
\usepackage{braket}

\usepackage{graphicx}
\usepackage{dcolumn}
\usepackage{bm}
\usepackage{xcolor}
\usepackage{hyperref}
\hypersetup{
    colorlinks=true,    
    linkcolor=blue,     
    citecolor=blue,     
    urlcolor=blue,      
    breaklinks=true     
}

\newcommand{\subfigimg}[3][,]{%
  \setbox1=\hbox{\includegraphics[#1]{#3}}%
  \leavevmode\rlap{\usebox1}%
  \rlap{\hspace*{-4pt}\raisebox{13.5\baselineskip}{#2}}%
  \phantom{\usebox1}%
}

\newcommand{\subfigimgtwo}[3][,]{%
  \setbox1=\hbox{\includegraphics[#1]{#3}}%
  \leavevmode\rlap{\usebox1}%
  \rlap{\hspace*{-4pt}\raisebox{14.4\baselineskip}{#2}}%
  \phantom{\usebox1}%
}

\begin{document}

\preprint{APS/123-QED}

\title{Complex quantum momentum due to correlation}

\author{Matthew Albert}
\altaffiliation{malbe058@uottawa.ca}

\author{Xiaoyi Bao}
\altaffiliation{xiaoyi.bao@uottawa.ca}
\affiliation{Department of Physics, University of Ottawa, Ottawa K1N 6N5, Canada}

\author{Liang Chen}
\altaffiliation{liang.chen@uottawa.ca}
\affiliation{Department of Physics, University of Ottawa, Ottawa K1N 6N5, Canada}
\date{\today}

\begin{abstract}
Real numbers provide a sufficient description of classical physics and all measurable phenomena; however, complex numbers are occasionally utilized as a convenient mathematical tool to aid our calculations. On the other hand, the formalism of quantum mechanics integrates complex numbers within its fundamental principles, and whether this arises out of necessity or not is an important question that many have attempted to answer. Here, we will consider two electrons in a one-dimensional quantum well where the interaction potential between the two electrons is attractive as opposed to the usual repulsive coulomb potential. Pairs of electrons exhibiting such effective attraction towards each other occur in other settings, namely within superconductivity. We will demonstrate that this attractive interaction leads to the necessity of complex momentum solutions, which further emphasizes the significance of complex numbers in quantum theory. The complex momentum solutions are solved using a perturbative analysis approach in tandem with Newton's method. The probability densities arising from these complex momentum solutions allow for a comparison with the probability densities of the typical real momentum solutions occurring from the standard repulsive interaction potential. 
\end{abstract}

\maketitle
\section{\label{sec:level1}Introduction}
\begin{figure}[ht]
\includegraphics[width=9cm]{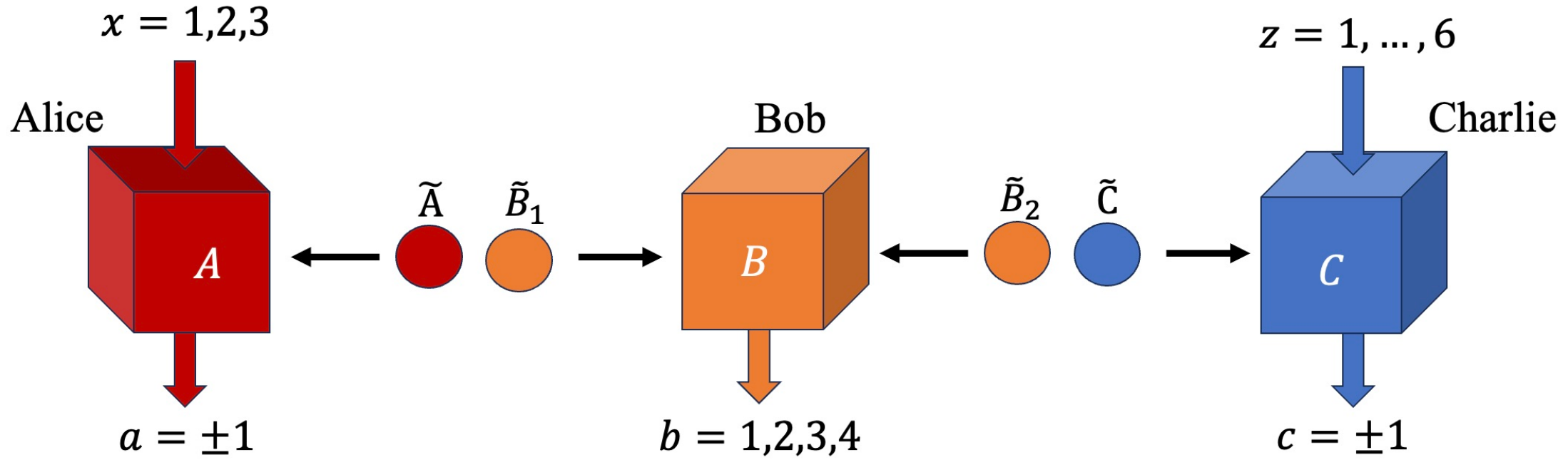}
\caption{\label{fig:experiment}Experimental setup proposed by Renou et al., two pairs of qubits are generated at different sources and distributed among three observers. The qubit pair $\Tilde{A}$ and $\Tilde{B}_1 (\Tilde{B}_2$ and $\Tilde{C})$ are sent to Alice and Bob (Bob and Charlie). After Bob performs a joint measurement on the qubits $\Tilde{B}_1$ and $\Tilde{B}_2$, the entanglement shifts to qubits $\Tilde{A}$ and $\Tilde{C}$. Alice and Bob select from three and six measurements to perform on qubits $\Tilde{A}$ and $\Tilde{C}$ respectively. (Figure adapted from Ref. \cite{renou2021quantum,MillerJohannaL.2022Dqmn}.)}
\end{figure}

 The prevalence of complex numbers throughout the quantum mechanic's formalism is apparent within its fundamentals. The Schr\"{o}dinger equation, wave functions, quantum operators, and commutation relations, to name a few, all employ its use extensively \cite{ChenMing-Cheng2022RORS}. Although complex numbers are a useful mathematical tool, physical meaning is attributed solely to real quantities (e.g., real eigenvalues to represent an observable and modulus squared of the wave function for probability), while complex quantities are thought of as non-physical \cite{KaramRicardo2020Wacn}. Whether complex numbers are necessary for quantum mechanics or if real numbers alone provide a complete description is a topic that many have explored \cite{MillerJohannaL.2022Dqmn,ChenMing-Cheng2022RORS,KaramRicardo2020Wacn,renou2021quantum, li2022testing,aleksandrova2013real,baez2012division,myrheim1999quantum}. Beginning with von Neumann in 1936 \cite{ChenMing-Cheng2022RORS,BednorzAdam2022Odbr,2c73be7a-3de4-3824-8e11-49ebe4b183e4}, numerous works have demonstrated the ability to simulate quantum systems based only on real numbers using an extended Hilbert space \cite{stueckelberg1960quantum,guenin1961quantum,McKagueMatthew2009Sqsu}. Conversely, in 2021, Renou et al. designed a Bell-like experiment that is potentially unexplainable by real quantum theory, hence providing compelling evidence towards the requirement of complex quantum theory for a complete explanation \cite{renou2021quantum}. First, defining quantum theory upon four postulates \cite{sep-qm,piron1964axiomatique}, the entanglement-swapping \cite{ZUKOWSKIM1993EBev} experiment comprised of three observers: Alice, Bob, and Charlie, as shown in Fig.~\hyperref[fig:experiment]{1}. Two pairs of maximally entangled qubits $(\tilde{A}$ and $\Tilde{B}_1,$ as well as $\Tilde{B_2}$ and $\Tilde{C})$ are generated from different sources. The first qubit pair $\Tilde{A}$ and $\Tilde{B}_1$ are sent to Alice and Bob, respectively, while the second pair $\Tilde{B}_2$ and $\Tilde{C}$ are given to Bob and Charlie, respectively. A spin $\frac{1}{2}$ qubit consists of a two-dimensional complex vector \cite{MillerJohannaL.2022Dqmn}, which means that when Bob performs a joint measurement on the qubits $\Tilde{B}_1$ and $\Tilde{B}_2$, there are four possible outcomes. Once Bob performs this measurement, the entanglement between the pairs of qubits shifts to qubits $\Tilde{A}$ and $\Tilde{C}.$ The new entangled qubit state between Alice and Charlie must maximally violate the CHSH$_3$ inequality with a value of $6\sqrt{2}$ \cite{renou2021quantum}, which is an amalgamation of three CHSH inequalities \cite{AcínAntonio2015Orcf,BowlesJoseph2018SoPo}. Shortly after, two groups performed the proposed experiment observing results that supported the necessity of complex quantum theory \cite{li2022testing,ChenMing-Cheng2022RORS}. This work will further exemplify the indispensability of complex numbers in quantum mechanics, particularly the need for complex momentum solutions when considering two electrons confined in a one-dimensional quantum well with an attractive effective interaction potential between the two electrons. A first exposure to quantum mechanics typically starts with the problem of a single electron confined within a one-dimensional quantum well of length $L$, in which the momentum solutions are found to be $k= \frac{n\pi}{L}$ and $n\in \mathbb{N}^+$ \cite{alma9953490279605154}. The consideration of two interacting electrons usually involves a repulsive interaction potential resulting in real momentum solutions. However, other works have considered complex momentum solutions, but in the context of two free bosonic particles \cite{franchini2011notes} or within the one-dimensional Heisenberg model \cite{karlebethe,bethe1931theorie}.  
\section{Model and analysis}
Our model consists of two identical electrons, each with mass $m$, confined in a one-dimensional quantum well of length $L$. The Hamiltonian used to describe these two electrons within the well is given by:
\begin{equation}
    H = -\frac{\hbar^2}{2m}\frac{\partial^2}{\partial x_1^2} + V(x_1) -\frac{\hbar^2}{2m}\frac{\partial^2}{\partial x_2^2} + V(x_2) + \frac{\hbar^2}{2mL}U\delta(x_1-x_2),
    \label{eq:Hamiltonian}
\end{equation}
where $x_1$ and $x_2$ are the position vectors of the two electrons and $V(x_i) =  \left\{ \begin{array}{ll} 0, \quad 0 < x_i < L \\ \infty, \quad \text{otherwise} \end{array} \right.$. The interaction potential between the two electrons was chosen to be pointlike, taking the form of a $\delta-$function, also known as the shape-independent approximation \cite{busch1998two}. The $\delta-$function model was introduced by Frost \cite{WeberGunterG.1964DMfI} and it allows us to avoid any complications of the Coulomb potential, while still being able to emphasize the necessity of complex momentum solutions. If we were to consider a spin triplet state with an antisymmetrical spatial factor of the wave function, this would lead to the same wavefunction that comes about from the non-interacting case and is given by:
\begin{equation}
    \begin{split}
        \Psi_{k_1,k_2}(x_1,x_2) = \frac{\sqrt{2}}{L}[\sin(k_1 x_1)\sin(k_2 x_2) \\
        - \sin(k_2 x_1)\sin(k_1 x_2)],
    \end{split}
\end{equation}
where $k_1 L= n\pi$ and $k_2 L=m\pi$ and $n\neq m$. We can see that the delta interaction potential plays no role, and the momentum solutions remain real. As outlined in the Supplemental Material (SM) \cite{SupplementalMaterial}, we will now consider a singlet state formed by the two electron spins, the Pauli exclusion principle requires a symmetrical spatial factor of the wave function \cite{alma992188765705151}. Consequently, this leads to a wavefunction of the form:
\begin{widetext}
\begin{equation}
    \Psi_{k_1,k_2}(x_1,x_2) =  \left\{ \begin{array}{ll} N\sin(k_1 x_1)\sin(k_2(L-x_2))+M\sin(k_2x_1)\sin(k_1(L-x_2)), \quad \text{if } x_1 < x_2 \\ N\sin(k_1 x_2)\sin(k_2(L-x_1))+M\sin(k_2x_2)\sin(k_1(L-x_1)), \quad \text{if } x_1 >x_2, \end{array} \right. 
 \label{eq:wideeq}
\end{equation}
\end{widetext}
where $N$ and $M$ are normalization constants to be determined. The center-of-mass coordinate $\xi = \frac{1}{\sqrt{2}}(x_1+x_2)$ and the relative coordinate $\eta = \frac{1}{\sqrt{2}}(x_1-x_2)$ \cite{busch1998two} allow the Hamiltonian to be rewritten as: 
\begin{equation}
    H = -\frac{\hbar^2}{2m}\left(\frac{\partial^2}{\partial \xi^2} +\frac{\partial^2}{\partial \eta^2}\right)+ V(x_1) + V(x_2) + \frac{\hbar^2}{2mL}U\delta(\sqrt{2}\eta).
\end{equation}
In the region $0<x_1,x_2 <L,$ the Schr\"{o}dinger equation gives: 
\begin{equation}
    \left[-\frac{\hbar^2}{2m}\left(\frac{\partial^2}{\partial \xi^2} +\frac{\partial^2}{\partial \eta^2}\right)+\frac{\hbar^2}{2mL}U\delta(\sqrt{2}\eta) \right] \Psi_{k_1,k_2} = E  \Psi_{k_1,k_2}.
\end{equation}
In the usual way, we can integrate the Schr\"{o}dinger equation from $-\epsilon$ to $\epsilon$ in the limit $\epsilon \rightarrow 0$ \cite{WeberGunterG.1964DMfI}
\begin{equation}
\begin{split}
&-\frac{\hbar^2}{2m}\int_{-\epsilon}^{\epsilon}\left(\frac{\partial^2\Psi_{k_1,k_2}}{\partial \xi^2} + \frac{\partial^2 \Psi_{k_1,k_2}}{\partial \eta^2}\right) d\eta \\
&\quad + \frac{\hbar^2}{2mL}U\int_{-\epsilon}^{\epsilon}\delta(\sqrt{2}\eta) \Psi_{k_1,k_2}d\eta = \int_{-\epsilon}^{\epsilon} E  \Psi_{k_1,k_2}(x_1,x_2)d\eta.
\end{split}
\end{equation}
After some algebra and redefining $k_1$ and $k_2$ to be dimensionless quantities, we obtain two cases, in which each case has two corresponding equations that the solutions $k_1$ and $k_2$ must satisfy. The first case is when $\frac{N}{M}=1,$ which has the two equations:
\begin{equation}
    \left\{ \begin{array}{ll} \frac{k_1 \sin(k_2 )}{k_2 \sin(k_1)}=-1, \\ 2(k_2\cot(k_2)+k_1\cot(k_1))= -U. \end{array} \right.
    \label{eq:system1}
\end{equation}
The second case where $\frac{N}{M}=-1$, which gives the two equations:
\begin{equation}
    \left\{ \begin{array}{ll} \frac{k_1 \sin(k_2)}{k_2 \sin(k_1)}=1, \\ 2(k_1\cot(k_1)+k_2\cot(k_2))= -U. \end{array} \right.
    \label{eq:system2}
\end{equation}
A comparison with the non-interacting case leads one to expect $k_1$ and $k_2$ solutions of Eq.~(\hyperref[eq:system1]{7}) and (\hyperref[eq:system2]{8}) to lie in the neighborhood $k_1 = n \pi$ and $k_2 = m \pi$ where $n$ and $m$ are positive integers. Momentum solutions satisfying Eq.~(\hyperref[eq:system1]{7}) require $n$ and $m$ to be of the same parity, while those satisfying Eq.~(\hyperref[eq:system2]{8}) are required to be of different parity. If we let $x = k_1$ and $y = k_2$ and perform a perturbative analysis (see the SM \cite{SupplementalMaterial}) around the region where $n=m $ that is $x = n\pi + \delta_x$ and $y = n\pi + \delta_y,$ we obtain: 

\begin{equation}
    \delta_x=\frac{U}{2n\pi}\pm \left[\left(\frac{U}{2n\pi}\right)^2+\frac{U-\frac{U^3}{6n^2\pi^2}}{2+\frac{U}{2}}\right]^{\frac{1}{2}},
    \label{eq:deltax}
\end{equation}
\begin{equation}
    \delta_y=\frac{U}{2n\pi}\mp \left[\left(\frac{U}{2n\pi}\right)^2+\frac{U-\frac{U^3}{6n^2\pi^2}}{2+\frac{U}{2}}\right]^{\frac{1}{2}}.
    \label{eq:deltay}
\end{equation}
If $U>0,$ the interaction potential between the two electrons is repulsive and we expect to find solutions where $k_1$ and $k_2$ are real. However, it is more interesting if we consider $U<0,$ the interaction potential will be attractive and we expect the two electrons to have some binding tendencies. The effective attraction between two electrons has been seen within the context of superconductivity where in a metal, two electrons are attracted to each other forming a Cooper pair. Bardeen, Cooper, and Schrieffer explained this phenomenon by the interaction between electrons and the ion lattice  \cite{WeisskopfVictorF.1981Tfoc}. In this case, and as is the case within superconductivity, this attraction leads to the formation of a bound state \cite{PogosovWalterV.2010Tpat}. It is important to notice that when $U<0,$ the radicand in Eq.~(\hyperref[eq:deltax]{9}) and (\hyperref[eq:deltay]{10}) can become negative resulting in complex solutions for $k_1$ and $k_2.$ We can use $k_1=n\pi+\delta_x$ and $k_2 = n\pi + \delta_y$ as our initial guess with Newton's method to obtain our exact solutions \cite{alma991028269309705161}. 

\section{Results and discussion}
\begin{figure*}
  \centering
  \begin{tabular}{@{}p{0.45\linewidth}@{\quad}p{0.445\linewidth}@{}}
    \subfigimg[width=\linewidth]{\textbf{(a)}}{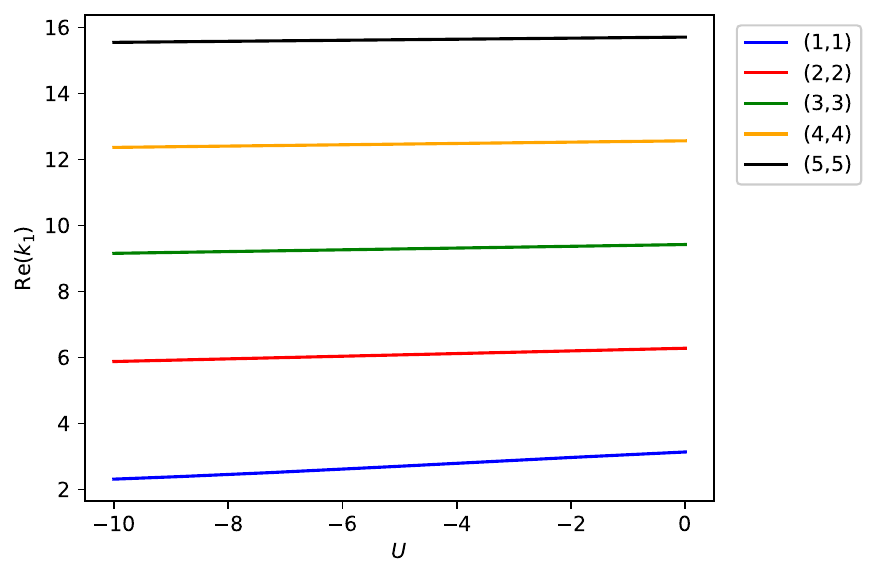} &
    \subfigimg[width=\linewidth]{\textbf{(b)}}{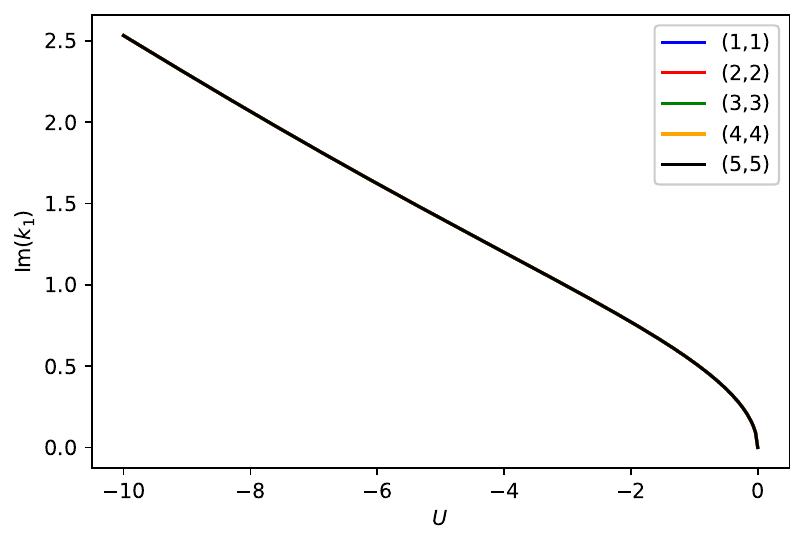} \\
    \subfigimg[width=\linewidth]{\textbf{(c)}}{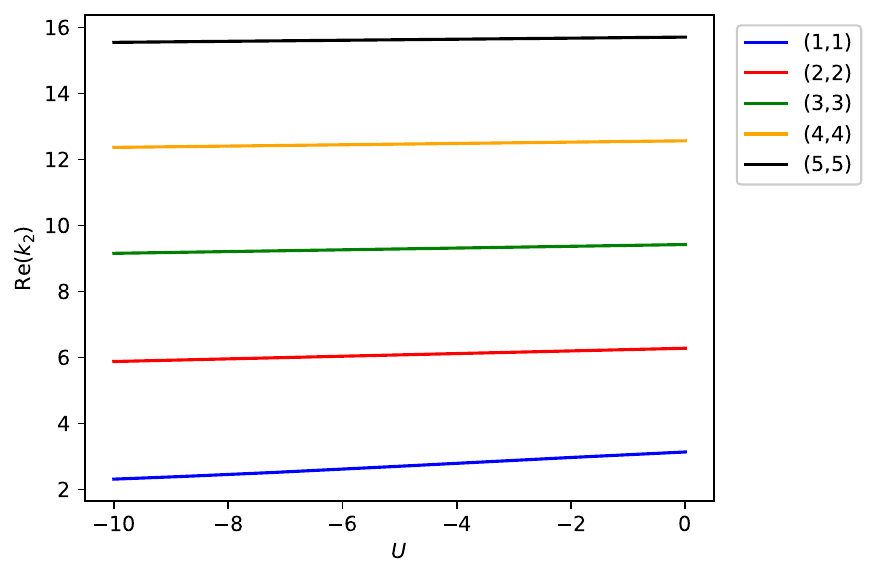} &
    \subfigimg[width=\linewidth]{\textbf{(d)}}{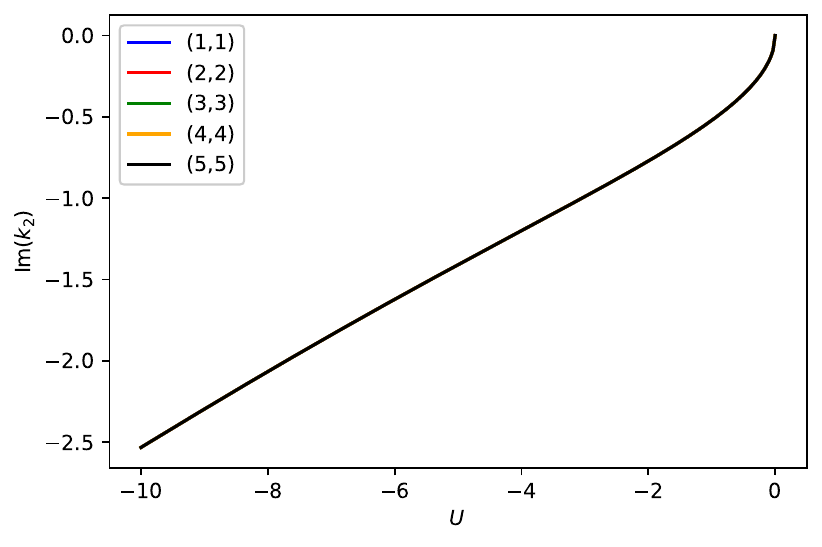}
  \end{tabular}
  \caption{\label{fig:momentum}Real and imaginary components of complex momentum solutions ($k_1$ and $k_2$) as a function of negative interaction potential (ranging from $-10$ to $0$) for identical parity. Although the imaginary components for the states considered are indistinguishable in (b) and (d), they exhibit very minute variations that are not observable in the plots (more so for $|U| << 1$).}
\end{figure*}
\begin{figure*}
  \centering
  \begin{tabular}{@{}p{0.45\linewidth}@{\quad}p{0.44\linewidth}@{}}
    \subfigimgtwo[width=\linewidth]{\textbf{(a)}}{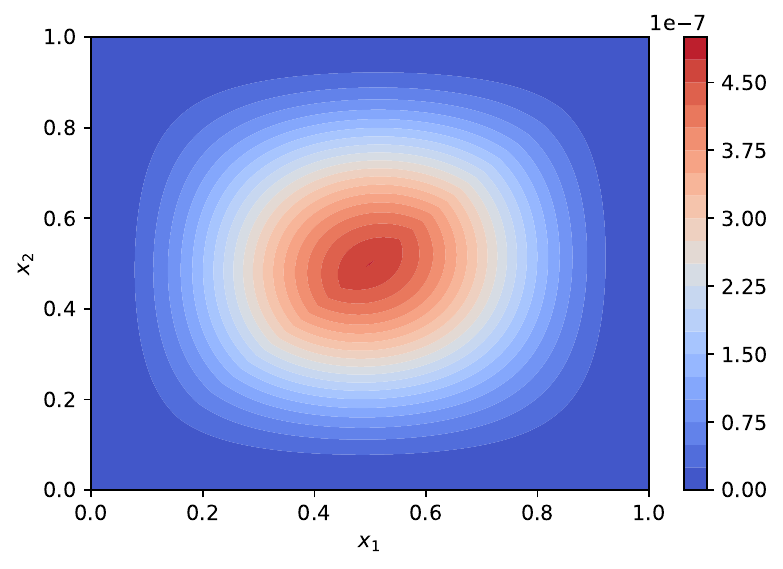} &
    \subfigimgtwo[width=\linewidth]{\textbf{(b)}}{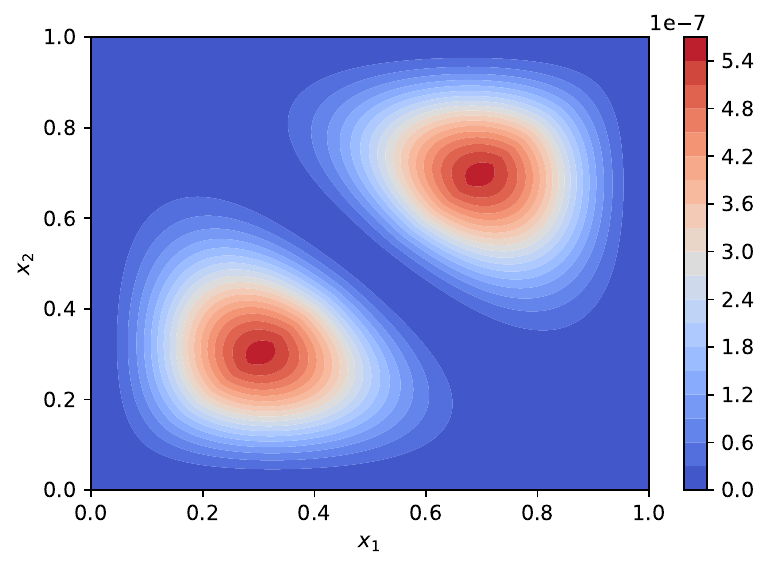} \\
    \subfigimgtwo[width=\linewidth]{\textbf{(c)}}{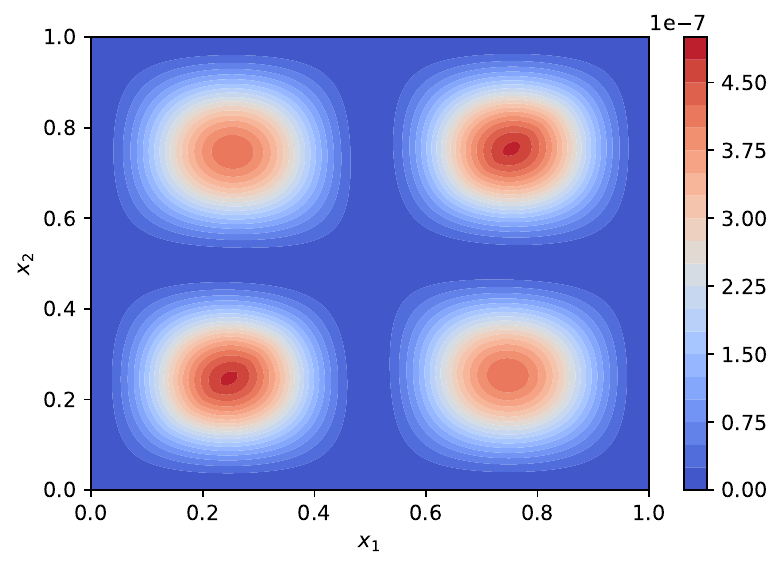} &
    \subfigimgtwo[width=\linewidth]{\textbf{(d)}}{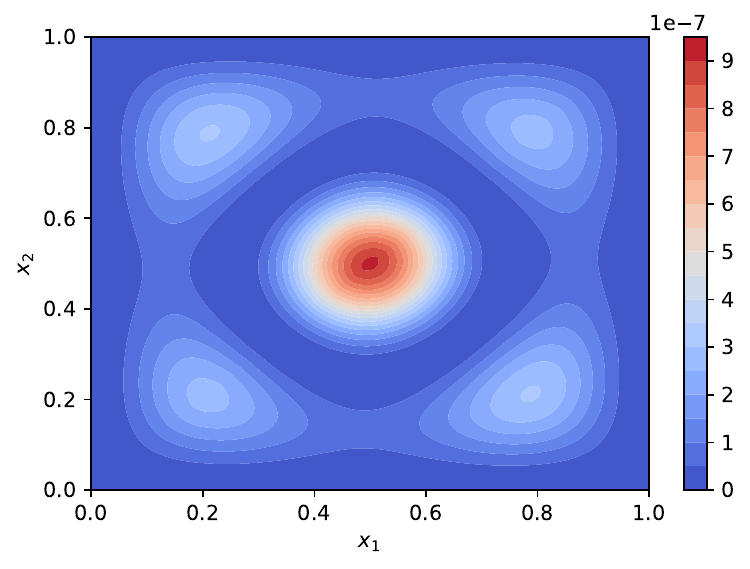}
  \end{tabular}
  \caption{\label{fig:wavefunction}Filled contours of the probability density for the four lowest energy states with a negative interaction potential of $U=-1$ are shown, where the positions of the two electrons, $x_1$ and $x_2$ are scaled with respect to $L$. (a) Complex solutions are given by 
  $k_1 = 3.06+0.52i$ and $k_2 = 3.06-0.52i$ corresponding to $(1,1)$ state. (b) Real momentum solutions are given by 
  $k_1 =6.05$ and $k_2 = 3.27$ corresponding to $(2,1)$ state.
  (c)  Complex solutions
  $k_1 = 6.24+0.52i$ and $k_2 = 6.24-0.52i$ corresponding to $(2,2)$ state. (d) Real momentum solutions
  $k_1 =9.30$ and $k_2 = 3.18$ corresponding to $(3,1)$ state.}
\end{figure*}
\begin{figure*}
  \centering
  \begin{tabular}{@{}p{0.45\linewidth}@{\quad}p{0.45\linewidth}@{}}
    \subfigimgtwo[width=\linewidth]{\textbf{(a)}}{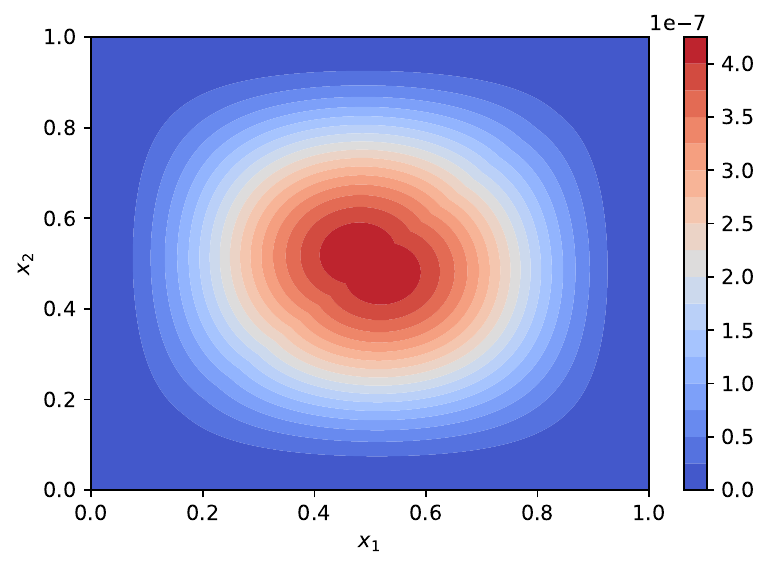} &
    \subfigimgtwo[width=\linewidth]{\textbf{(b)}}{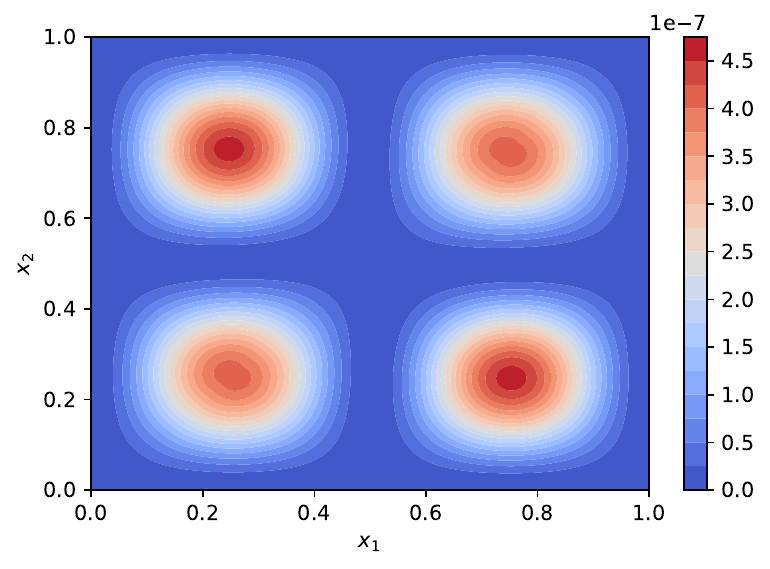}
  \end{tabular}
  \caption{\label{fig:real}Filled contours of the probability density for two identical parity states with a positive interaction potential of $U=1$ are shown, where the positions of the two electrons, $x_1$ and $x_2$ are scaled with respect to $L$.  (a) Real momentum solutions are given by 
  $k_1 =3.70$ and $k_2 = 2.74$ corresponding to $(1,1)$ state. (b) Real momentum solutions are given by 
  $k_1 = 6.80$ and $k_2 = 5.84$ corresponding to $(2,2)$ state.}
\end{figure*}

We plotted the complex momentum solutions for identical parity (i.e., when $n=m$) as a function of a negative interaction potential in Fig.~\hyperref[fig:momentum]{2}. In particular, the complex solutions $k_1$ and $k_2$ are complex conjugates of one another, which ensures that the energy remains real. The imaginary components for the different parity, i.e., $(1,1),...,(5,5)$ are essentially identical up to a sufficient level of numerical accuracy as illustrated in Fig.~\hyperref[fig:momentum]{2(b)} and \hyperref[fig:momentum]{2(d)}, although there are very slight differences in their values that are not discernable in the plots. These very minute differences arise from the weak dependence on $n$ in the radicand of Eq.~(\hyperref[eq:deltax]{9}) and (\hyperref[eq:deltay]{10}), and the magnitude of their differences are more noticeable for $|U|<<1$. It was found that this perturbative treatment, combined with Newton's method, does not yield correct momentum solutions for non-identical parity (i.e., when $n\neq m$). Consequently, the configuration interaction (CI) method was utilized to initially obtain a reliable approximation for the energy. This effectively reduces the degrees of freedom in our search, enabling Newton's method to be implemented successfully (see the SM \cite{SupplementalMaterial}). The momentum solutions for non-identical parity are purely real, whereas complex momentum solutions occur only for identical parity. The reason is that momentum solutions where $n \neq m$ cannot be complex conjugates of each other. The solutions must satisfy either Eq.~(\hyperref[eq:system1]{7}) or Eq.~(\hyperref[eq:system2]{8}), with the additional constraint of real energy given by $E=\frac{\hbar^2}{2mL^2}(k_1^2+k_2^2)$, which can only be simultaneously fulfilled by real momentum solutions when parity $n\neq m$. With the momentum solutions obtained, we can utilize Eq.~(\hyperref[eq:wideeq]{3}) to plot the probability density $|\Psi_{k_1,k_2}(x_1,x_2)|^2$ for different states with a negative interaction potential of $U=-1$, as depicted in Fig.~\hyperref[fig:wavefunction]{3}. The probability densities in Fig.~\hyperref[fig:wavefunction]{3(a)} and \hyperref[fig:wavefunction]{3(c)} involve complex momentum solutions, whereas those in Fig.~\hyperref[fig:wavefunction]{3(b)} and \hyperref[fig:wavefunction]{3(d)} are from real momentum solutions. We can compare the wavefunction under negative and positive interaction potentials by changing the value of $U$ from $-1$ to $1$ and plotting the resulting probability density in Fig.~\hyperref[fig:real]{4}. In Fig.~\hyperref[fig:wavefunction]{3(c)}, the probability density corresponding to the $(2,2)$ state is weaker along the $x_1+x_2=1$ diagonal and stronger along the $x_1=x_2$ diagonal. This is because, along the $x_1+x_2=1$ diagonal, the electrons are maximally separated from each other within the 1D quantum well. As a result, the attractive delta-function interaction plays no role. Conversely, along the $x_1=x_2$ diagonal, the two electrons are in the same position, and the negative interaction potential increases the probability of finding the electrons in this region. Fig.~\hyperref[fig:real]{4(b)} displays a similar behavior but in reverse for a repulsive positive interaction potential where the probability density corresponding to the $(2,2)$ state is stronger along the $x_1+x_2=1$ diagonal and weaker along the $x_1=x_2$ diagonal. 

\section{Conclusions}
In summary, we have demonstrated that two electrons forming a singlet state confined in a one-dimensional infinite quantum well, modeled by an effective attractive delta-function interaction potential between the two electrons, require complex momentum solutions in the case of identical parity. Perturbative analysis and Newton's method allowed us to obtain these complex momentum solutions, which consisted of conjugate pairs. As a result, this provides evidence for the necessity of complex numbers within quantum theory while considering a simplified model. Although a more complicated model involving many-body effects from more than two electrons, a different attractive interaction potential, and higher dimensionality exists, our simplified model sufficiently highlights the importance of complex numbers in quantum mechanics. Furthermore, by incorporating the configuration interaction method, we also acquired real momentum solutions for non-identical parity. The probability densities for positive and negative interaction potentials appeared similar but differed most noticeably along the diagonals, i.e., when the two electrons had the same position, the negative(positive) interaction potential increased(decreased) the probability densities.
\begin{acknowledgments}
Matthew Albert would like to acknowledge Daniel Miravet for the fruitful discussions on numerical implementations and for providing valuable resources. Liang Chen would like to acknowledge the discussion with Vladimir Kalosha. 
\end{acknowledgments}

%

\section{Supplemental Material}
\subsection{Singlet Wavefunction}
In this appendix, we will begin by deriving the expression for the singlet state as seen in Section II of the main text. First, consider a singlet spin state formed by the two electron spins. The Pauli exclusion principle demands that the spatial wave function be symmetric under the permutation operation $P_{12}$ \cite{alma992188765705151}. Formally we could write:
\begin{equation}
\ket{k_1,k_2}=\ket{1:k_1,2:k_2}+\ket{1:k_2,2:k_1},
\end{equation}
and written spatially as: 
\begin{equation}
    \ket{x_1,x_2}=\ket{1:x_1,2:x_2}+\ket{1:x_2,2:x_1}. 
\end{equation}
Projection to spatial wave function gives:
\begin{equation}
    \begin{split}
        \braket{x_1,x_2|k_1,k_2} =\braket{x_1,x_2|k_1,k_2}+\braket{x_1,x_2|k_2,k_1}\\
        +\braket{x_2,x_1|k_1,k_2}+\braket{x_2,x_1|k_2,k_1}.
    \end{split}
\end{equation}
Which can be expanded as:
\begin{widetext}
\begin{equation}
   \Psi_{k_1,k_2}(x_1,x_2)=  \left\{ \begin{array}{ll} 2A_{k_1}B_{k_2}\sin(k_1 x_1)\sin(k_2(L-x_2))+2A_{k_2}B_{k_1}\sin(k_2x_1)\sin(k_1(L-x_2)), \quad \text{if } x_1 < x_2 \\ 2A_{k_2}B_{k_1}\sin(k_2x_2)\sin(k_1(L-x_1))+2A_{k_1}B_{k_2}\sin(k_1 x_2)\sin(k_2(L-x_1)), \quad \text{if } x_1 >x_2, \end{array} \right.
 \label{eq:wideeq2}
\end{equation}
\end{widetext}
where defining $N=2A_{k_1}B_{k_2}$ and $M=2A_{k_2}B_{k_1}$ as normalization constants give Eq.~(\hyperref[eq:wideeq]{3}) in the main text. 
\subsection{Perturbative Analysis}
As stated in the main text, a comparison with the non-interacting case leads one to expect $k_1$ and $k_2$ solutions of Eq.~(\hyperref[eq:system1]{7}) and (\hyperref[eq:system2]{8}) (in the main text) to lie in the neighborhood $k_1 = n \pi$ and $k_2 = m \pi$ where $n,m \in \mathbb{N}$. We can let $x = k_1$ and $y = k_2,$ then Eq.~(\hyperref[eq:system1]{7}) becomes:
\begin{equation}
    \left\{ \begin{array}{ll} x\sin(y)+y\sin(x)=0, 
\\ 2(x\cot(x)+y\cot(y))=-U. \end{array} \right.
\label{eq:system11}
\end{equation}
Consider a small deviation $\delta_x$ in $x$ and $\delta_y$ in $y,$ that is assume $x = n\pi + \delta_x \text{ and } y = m\pi + \delta_y.$ Making this substitution along with Taylor approximating $\sin(\delta_y) \approx \delta_y - \frac{1}{3!}\delta_y^3$ and keeping only terms up to third order in $\delta_x, \delta_y$ from the first of Eq.~(\hyperref[eq:system11]{15}) gives:
\begin{equation}
    \pi(n\delta_y + m\delta_x)+2\delta_x\delta_y-\frac{\pi}{3!}(n\delta_y^3 + m\delta_x^3)=0.
    \label{eq:system22}
\end{equation}
Similarly for the second of Eq.~(\hyperref[eq:system11]{15}) 
and using Eq.~(\hyperref[eq:system22]{16}) gives:
\begin{equation}
    n\delta_x + m \delta_y= \frac{U}{\pi}.
\end{equation}
Solving for $\delta_x$ gives:
\begin{equation}
        \delta_x = \frac{1}{n}(\frac{U}{\pi}-m\delta_y),
\end{equation}
which can be substituted into Eq.~(\hyperref[eq:system22]{16}) for the case where $n=m$ yielding:
\begin{equation}
    (U-\frac{U^3}{6n^2\pi^2})+(\frac{2U}{n\pi}+\frac{U^2}{2n\pi})\delta_y+(-2-\frac{U}{2})\delta_y^2=0.
    \label{eq:system33}
\end{equation}
Using the quadratic formula on Eq.~(\hyperref[eq:system33]{19}) allows us to obtain Eq.~(\hyperref[eq:deltax]{9}) and (\hyperref[eq:deltay]{10}) in the main text. The same analysis can be done on Eq.~(\hyperref[eq:system2]{8}) of the main text, and it will result in the same expressions for the case of $n=m$. 

\subsection{Configuration Interaction Method}
Defining $x_1=L\xi$ and $x_2=L\eta,$ the Hamiltonian in Eq.~(\hyperref[eq:Hamiltonian]{1}) of the main text, in the region $0<x_1,x_2 < L$ can be written as:
\begin{equation}
     H = -\frac{\hbar^2}{2mL^2}\frac{\partial^2}{\partial \xi^2} -\frac{\hbar^2}{2mL^2}\frac{\partial^2}{\partial \eta^2} + \frac{\hbar^2}{2mL^2}U\delta(\xi-\eta).
    \label{eq:system4}
\end{equation}
We can then scale the Hamiltonian by $\frac{\hbar^2}{2mL^2}$ to obtain: 
\begin{equation}
     \Tilde{H} = -\frac{\partial^2}{\partial \xi^2} -\frac{\partial^2}{\partial \eta^2} + U\delta(\xi-\eta). 
    \label{eq:system5}
\end{equation}
The normalized non-interacting single particle eigenstates are given by $\psi_n(x) = \sqrt{2} \sin(n\pi x),$ with energy $E_n = n^2\pi^2$ and they satisfy the condition $\int \psi_n^*(x) \psi_m(x) = \delta_{n,m}.$ The normalized symmetric wavefunction is given by:
\begin{equation}
    \ket{n,m} = N_{n,m} (\psi_n(\xi) \psi_m(\eta) + \psi_m(\xi) \psi_n(\eta)), 
\end{equation}
where $N_{n,m} =  \left\{ \begin{array}{ll} \frac{1}{2}, \quad \text{if } n=m \\ \frac{1}{\sqrt{2}}, \quad \text{if } n \neq m \end{array} \right.$. First we solve for the non-interacting elements of our Hamiltonian, this yields:
\begin{widetext}
\begin{equation}
    \braket{n,m|-\frac{\partial^2}{\partial \xi^2}-\frac{\partial^2}{\partial \eta^2} |\Tilde{n},\Tilde{m}} =2\pi^2(\Tilde{n}^2+\Tilde{m}^2) N_{\Tilde{n},\Tilde{m}}N_{n,m} (\delta_{n,\Tilde{n}}\delta_{m,\Tilde{m}}+\delta_{n,\Tilde{m}}\delta_{\Tilde{n},m}).
    \label{eq:system6}
\end{equation}
\end{widetext}
Similarly a longer expression for the interacting elements of the Hamiltonian can also be obtained:
\begin{widetext}
\begin{equation}
    \braket{n,m|U\delta(\xi-\eta)|\Tilde{n},\Tilde{m}} =2UN_{\Tilde{n},\Tilde{m}} N_{n,m}(\delta_{n+\Tilde{m},m+\Tilde{n}}+\delta_{n+\Tilde{n},m+\Tilde{m}}-\delta_{m+\Tilde{n}+\Tilde{m},n}-\delta_{n+\Tilde{n}+\Tilde{m},m}-\delta_{n+m+\Tilde{m},\Tilde{n}}-\delta_{n+m+\Tilde{n},\Tilde{m}}+\delta_{n+m,\Tilde{n}+\Tilde{m}}).
    \label{eq:system7}
\end{equation}
\end{widetext}
We can construct our Hamiltonian matrix using Eq.~(\hyperref[eq:system6]{23}) and (\hyperref[eq:system7]{24}), and then diagonalize this matrix to obtain the eigenenergies. Once we have the value of a particular energy $\Tilde{E}$ corresponding to a specific eigenstate, we can use the relation $\Tilde{E}=k_1^2+k_2^2$ (we have scaled the energies by $\frac{\hbar^2}{2mL^2}$). We can express $k_1$ and $k_2$ as:
\begin{equation}
    \left\{ \begin{array}{ll} k_1=\Omega \sin(\theta)+i\rho \cos(\theta), \\k_2=\Omega \cos(\theta)-i\rho \sin(\theta), \end{array} \right.
    \label{eq:solutions}
\end{equation}
where $\Omega,\rho, \theta \in \mathbb{R}$ and $\Omega, \rho >0.$ This choice ensures that the energy remains real, as $\Tilde{E}= \Omega^2-\rho^2.$ Since $\Omega=\sqrt{\Tilde{E}+\rho^2},$ we can rewrite Eq.~(\hyperref[eq:solutions]{25}):
\begin{equation}
    \left\{ \begin{array}{ll} k_1=\sqrt{\Tilde{E}+\rho^2} \sin(\theta)+i\rho \cos(\theta), \\k_2=\sqrt{\Tilde{E}+\rho^2} \cos(\theta)-i\rho \sin(\theta). \end{array} \right.
    \label{eq:solutions2}
\end{equation}
The number of free parameters has reduced from three in Eq.~(\hyperref[eq:solutions]{25}) to two in Eq.~(\hyperref[eq:solutions2]{26}), which allows for effective utilization of Newton's method to solve for these momentum solutions.
\maketitle


\end{document}